\documentstyle[prl,aps,twocolumn,epsf]{revtex} 
\def\break#1{\pagebreak \vspace*{#1}}
\def\epsfig#1#2#3#4
         {
         \epsfysize=#2 \vbox{ \hglue#3 \epsfbox[#4]{#1} }
         }
\def\epsfigrot#1#2#3#4
         {
         \epsfxsize=#2 
         \setbox\rotbox=\hbox to #2{\epsfbox[#4]{#1}}
         \vbox{\hglue#3 \rotl\rotbox}
         }
\newbox\rotbox
\input rotate
\begin{document}
\draft
\title{Friedel oscillations for interacting fermions in one dimension}
 \author{Reinhold Egger and Hermann Grabert}
\address{Fakult\"at f\"ur Physik, Albert-Ludwigs-Universit\"at,
Hermann-Herder-Stra{\ss}e 3, D-79104 Freiburg, Germany }
\maketitle
\widetext
\begin{abstract}
We study  Friedel oscillations in one-dimensional electron liquids
for arbitrary electron-electron interaction and 
arbitrary impurity strength.  
Explicit results for spinless as well as spin-$\frac12$
electrons are given.
In the case of Luttinger-liquid leads,
the Friedel oscillations decay as $x^{-g}$ far away from
the impurity where $g$ is the interaction constant. 
For a weak scatterer, a slower decay is found at small-to-intermediate
distances from the impurity, 
with a crossover to the asymptotic $x^{-g}$ decay.
\end{abstract}
\pacs{PACS numbers: 72.10.-d, 73.40.Gk}

\narrowtext

Quasi one-dimensional (1D) interacting fermion systems have 
attracted renewed attention recently, partly due to the 
technological relevance of such systems. They are also appealing because of 
 the existence of  exact solutions for simple models 
 \cite{luttinger,solyom,luther,emery,haldane,kane92,matveev1,wigner,fabrizio}.
The striking non-Fermi liquid behavior of 1D electrons 
is most clearly exhibited in transport properties,
especially in the presence of impurities or barriers\cite{kane92}.
Usually, either the case of a very weak or an almost insulating
barrier have been studied explicitly. The crossover between these two
regimes, however, has rarely been looked at despite its 
importance for several fundamental issues, e.g.~pinning of a Wigner
crystal \cite{pin} or charge density waves \cite{gruner}, 
breakdown of charging effects with increasing tunnel 
conductance \cite{sct}, or transport in 1D quantum wires 
or heterostructure channels \cite{altshuler}. In this Letter, we discuss 
several aspects  of this crossover, in particular properties
of the screening cloud around the impurity.

The presence of an impurity in a metal is known to cause
Friedel oscillations in the density profile due to the sharp 
Fermi surface \cite{friedel,zawa},
\[
\rho(r) \sim \cos(2k_{\rm F} r + \delta)/r^d\;,
\]
where $k_{\rm F}$ is the Fermi vector, $d$ the dimensionality, 
$r$ the distance from the impurity, and $\delta$ a phase shift.  
This result holds far away from the impurity and for a Fermi liquid
only, and in 1D the question arises how Friedel oscillations
are affected by the presence of strong electron correlations.
As pointed out by Matveev {\em et al.}\cite{matveev1},
for weakly interacting electrons in 1D, one can 
understand the zero-voltage anomaly and power-law temperature
dependence of the nonlinear conductance \cite{kane92}
in terms of electron backscattering
by Friedel oscillations. It is thus surprising how little attention
has been devoted to the modification of Friedel oscillations by
strong interactions.

We treat the 1D interacting electron liquid in the framework of standard
bosonization \cite{luther,emery,haldane,kane92}. This approach 
is appropriate for low temperatures where only excitations near
the Fermi surface are relevant. The electron 
creation operator for spin $s=\pm$ at position $x$ 
is expressed in terms of the 
boson fields $\theta_\mu(x)$ and $\phi_\mu(x)$ ($\mu=\rho,\sigma$)
\break{1.30in}
which arise as linear combinations of spin-up and spin-down fields,
\begin{eqnarray}\nonumber
\psi_s^\dagger(x) &\sim&  \sum_{n=\pm 1}
\exp \left[in\left(k_{\rm F} x+\sqrt{\pi/2} \,[\theta_\rho(x)+
s\theta_\sigma(x)]\right)
\right]\\&\times& \exp\left[ \label{field}
 i \sqrt{\pi/2} \,[\phi_\rho(x)+s \phi_\sigma(x) ] \right]\;.
\end{eqnarray}
The boson fields obey equal-time commutation relations
\[
[ \phi_\mu(x),\theta_\nu(x')] =  -(i/2) \delta_{\mu \nu} \mbox{sgn}(x-x')\;,
\]
and the canonical momentum  for the $\theta_\mu$ field is 
$\Pi_\mu  = \partial_x \phi_\mu$.  

We are concerned with density distributions in the presence of impurities or
barriers. The bosonized form of the density operator is \cite{wigner}
 \begin{eqnarray}
\hat{\rho}(x) &=& \sqrt{2/\pi}\, \partial_x \theta_\rho(x)
+ \frac{2k_{\rm F}}{\pi} \, \cos[2k_{\rm F}x+\sqrt{2\pi}\,\theta_\rho(x)]
 \nonumber    
\\    \label{dens}
&\times& \cos[\sqrt{2\pi}\,\theta_\sigma(x)]
+  {\rm const.} \cos[4k_{\rm F}x + \sqrt{8\pi}\, \theta_\rho(x)]\;,
\end{eqnarray}
where the background charge $\rho_0=2k_{\rm F}/\pi$ has been omitted.
The three terms in Eq.(\ref{dens})
are (1) the long-wavelength contribution, (2) the $2k_{\rm F}$ charge 
density wave part, and (3) the $4k_{\rm F}$ Wigner component 
\cite{emery,wigner}. The Wigner component is not present 
in the spinless case, since two right-movers have to be
flipped into left-movers simultaneously for this term to arise.

Assuming to be away from lattice or spin density wave
instabilities, and neglecting electron-electron 
backscattering for the moment,
the clean system is described by $H_0  = H_\sigma + H_\rho$ with
\begin{eqnarray}\label{hamil}
H_\rho&=& \frac{v_{\rm F}}{2} \int dx 
\left[\Pi_\rho^2(x) + (\partial_x \theta_\rho(x))^2\right]\\ \nonumber
 &+& \frac{1}{\pi} \int dx dx' \,U(x-x')\, \partial_x\theta_\rho(x)\,
\partial_{x'} \theta_\rho(x')\;,
\end{eqnarray}
where $v_{\rm F}$  is the Fermi velocity and $\hbar=1$.
The spin part $H_\sigma$ is identical to the charge part $H_\rho$
with no interaction potential $U$ and the $\rho$ fields replaced 
by the $\sigma$ fields. Here, $U(x)$ is a (screened) Coulomb interaction, 
and we will explicitly study a short-ranged
potential (Luttinger liquid) \cite{haldane} and a $1/r$
long-ranged potential \cite{wigner,fabrizio}. 

Let us now consider a scattering potential. Assuming 
an essentially pointlike scatterer at $x=0$, one finds
\begin{equation}\label{imp}
H_{\rm imp}= V  \cos[\sqrt{2\pi}\, \theta_\rho(0)] \, 
\cos[\sqrt{2\pi}\, \theta_\sigma(0)]\;.
\end{equation}
Spin and charge parts  are now coupled through this term\cite{kane92}.
Actual computations using the bosonized model $H_{\rm bos} = H_0+H_{\rm imp}$
necessitate introduction of a cutoff parameter $\omega_{\rm c}$\cite{foot}. 
Increasing $V/\omega_{\rm c}$ from zero to infinity
corresponds to tuning the barrier from transmittance 
one down to zero. Near zero transmittance, the 
weak-link model used in Ref.\cite{kane92} is reproduced 
by the instanton treatment of $H_{\rm bos}$.
One may also show \cite{prep} by direct comparison 
with the exactly solvable Fano-Anderson model
 that $H_{\rm bos}$ can reproduce 
the full crossover {\em quantitatively}, thus validating
Eq.(\ref{imp}) for arbitrary $V/\omega_{\rm c}$. Although this
comparison can be carried out only in the absence of Coulomb interactions,
it implies that a complete description of the crossover is indeed possible
using $H_{\rm bos}$.

Friedel oscillations can be extracted from the generating functional
\begin{equation}\label{gene}
Z(x,\lambda^{}_\mu) = \left\langle  \exp
\left[\sqrt{2\pi} \, i \sum_{\mu = \rho,\sigma} 
\lambda^{}_\mu \theta^{}_\mu(x)  
\right] \right \rangle\;,\nonumber
\end{equation}
where the average is taken over $H_{\rm bos}$.
Since the impurity influences $\theta_\mu$ only at $x=0$,
we constrain $\theta_\mu(x=0)$ to be equal to new fields,
say, $q_\mu =\sqrt{2\pi}\, \theta_\mu(0)$.  Representing these
constraints by a Fourier functional integral, one can integrate
out all $\theta_\mu(x)$ modes due to their Gaussian nature.  
In contrast to previous treatments of this problem, 
we keep explicit information about the electron liquid away from
the barrier. The remaining auxiliary field integrations coming from  the
above constraints are Gaussian and hence also performed easily. In the
end, one is left with the nontrivial average over the $q_\mu$ fields
alone, which are coupled to each other through $H_{\rm imp}$.

Collecting together all terms, we obtain
\begin{equation}\label{zres}
Z = {\cal B}(x,\lambda_\mu,V) \prod_{\mu=\rho,\sigma} W^{\lambda_\mu^2}_\mu(x)
\;.
\end{equation}
The functions $W_\mu(x)$
 are independent of the barrier height
since they  do not participate in the $q$ average,
\begin{equation}\label{ww}
W_\mu(x) = \exp\left(\frac{1}{\beta} \sum_{n=-\infty}^\infty 
\frac{F^{(\mu )\,2}_n(x)-F^{(\mu )\,2}_n(0)}{ F^{(\mu )}_n(0)}
\right)\;,
 \label{defx} 
\end{equation}
where $\beta=1/k_{\rm B}T$ and 
\begin{equation} \label{ff}
F^{(\mu )}_n(x)= v_{\rm F} \int_{-\infty}^\infty dk \,\frac{\cos (kx)}{
\omega_n^2  + v_{\rm F}^2 k^2 (1+2U_k\delta_{\mu\rho} /\pi v_{\rm F})}\;.
\end{equation}
Here, $\omega_n=2\pi n/\beta$ are the Matsubara frequencies
and $U_k$ is the Fourier transformed electron-electron
 interaction \cite{fot1}.
A Luttinger liquid is governed by the interaction 
constants $g_\sigma=1$ and $g_\rho = g \leq 1$ \cite{haldane,kane92}.
In that case, Eq.(\ref{ff}) becomes simply
\[
F^{(\mu)}_n(x) = \frac{\pi g_\mu}{|\omega_n|}
 \exp\left[-\frac{|g_\mu \omega_n x|}{v_{\rm F}}\right] \;.
\]
Finally, the  quantity ${\cal B}$ in Eq.(\ref{zres}) is 
 an average in $q$ space; all dependency
on impurity properties is contained in this factor. With Matsubara
components $q_{\mu,n}$, we find
\[
{\cal B}(x,\lambda_\mu,V) =\Biggl\langle
 \prod_\mu \exp\Biggl[
 \frac{i\lambda_\mu}{\beta} \sum_{n=-\infty}^\infty q_{\mu,n}
 \frac{F^{(\mu )}_n(x)}{F^{(\mu )}_n(0)}
\Biggr ]\Biggr\rangle_q\;. 
\]
The $q$ bracket stands for an average taking the action
\[
S[q_\mu]= \frac{1}{\beta} \sum_\mu \sum_{n=1}^\infty \frac{|q_{\mu,n}|^2}
{F^{(\mu)}_n(0)} + V\int_0^\beta d\tau \cos [q_\sigma(\tau)]
\cos [q_\rho(\tau)] \;.
\]
The first term is the standard influence functional \cite{kane92}. From
these equations [or generalizations with additional
$\theta$ fields at some other position $x'$], one may obtain all desired
information about density profiles and correlations in presence
of an arbitrarily high barrier. 
The results of Refs.\cite{haldane,wigner} for the clean system
($V=0$) are easily recovered. Similar expressions incorporating
the $\phi$ fields reproduce the results of Refs.\cite{kane92,fabrizio}.

Let us now consider the expectation value
$\langle \hat\rho(x) \rangle$. From Eq.(\ref{zres}), one finds $\langle 
\theta_\mu(x)\rangle=0$. The long-wavelength part in $\hat \rho$
[first term in Eq.(\ref{dens})]
does not feel the impurity since $H_{\rm imp}$ does not contain
forward scattering terms. However, Friedel oscillations follow for the
$2k_{\rm F}$ and (in the spin-$\frac12$ case)
for the $4k_{\rm F}$ component in Eq.(\ref{dens}).
 
For a spinless Luttinger liquid, we obtain the Friedel oscillation 
\begin{equation}\label{friedel}
\langle \hat \rho(x)\rangle /\rho_0 = - 
P(|x|,g,V) \, W(|x|,g) \,
\cos(2k_{\rm F} x) \;,
\end{equation}
where $\rho_0=k_{\rm F}/\pi$.
Evaluation of ${\cal B}$ gives the {\em pinning function} 
\begin{equation} \label{pinning}
P(x,g,V) = -
\left \langle \cos\left[\frac{1}{\beta}\sum_{n=-\infty}^\infty 
 e^{-gx|\omega_n| /v_{\rm F} } q_n \right] \right\rangle_q\;.
\end{equation}
This function determines the amplitude of the Friedel 
oscillation and hence the ability of the scatterer to pin 
charge density waves.

In the following, we discuss the ground-state properties of the Friedel
oscillation (\ref{friedel}) in some detail. From Eq.(\ref{ww}), we find
\begin{equation}
 W(x,g) = (1+x/\alpha)^{-g} \;,
\end{equation}
where $\alpha=v_{\rm F}/2 g \omega_{\rm c}$ is a microscopic lengthscale,
say, a lattice spacing.
The properties of the pinning function $P$ can be studied
using either numerically exact quantum Monte Carlo (QMC)
simulations or simple approximations.
For transmittance one ($V=0$), the ``charge'' $q$ is free and $P=0$.
For zero transmittance ($V\to \infty$), 
the potential $V \cos q$ locks $q$ at odd multiples of $\pi$, and $P$
takes its maximal value, $P=1$, for all $x$. 

To estimate $P$ for arbitrary $V/\omega_{\rm c}$, we
first discuss a simple variational procedure based on a quadratic 
trial Hamiltonian (self-consistent harmonic approximation, SCHA)
 \cite{var,gog1}. Replacing the cosine term by a Gaussian with 
frequency $\Omega$, Feynman's variational principle leads to the 
self-consistency equation
\begin{equation} \label{selfcon}
\Omega = V \left(1+\frac{\omega_{\rm c}}{2\pi g \Omega}\right)^{-g}\;.
\end{equation} 
Within the SCHA, Eq.(\ref{pinning}) is a Gaussian average, and one finds 
\begin{equation} \label{respin}
P(x,g,V) = \exp\left[ -g\, e^{(x+\alpha)/x_0}\, E_1((x+\alpha)/x_0) \right]
\end{equation}
with the exponential integral $E_1(y)$ \cite{abram}
and the crossover scale
\begin{equation} \label{cross}
x_0 = \frac{\alpha}{ 2\pi g} \frac{\omega_{\rm c}}{\Omega}\;.
\end{equation}

In the strong-scattering limit, $\pi V /\omega_{\rm c} \gg 1$,  
Eq.(\ref{selfcon}) yields $\Omega=V$. In this limit, only small 
fluctuations around the 
minima of the cosine potential are possible, with interwell tunneling  
being forbidden by an exponentially small WKB factor. 
Since $x_0$ is even smaller than $\alpha$, see Eq.(\ref{cross}), 
the term ``crossover'' is not meaningful in this limit.
Using asymptotic properties of  $E_1(y)$, Eq.(\ref{respin}) becomes 
for $x\gg \alpha$
\begin{equation} \label{Plarge}
P = e^{-g x_0/x} \simeq 1\;, 
\end{equation}
in accordance with our QMC results and a recent study of  Friedel 
oscillations by open boundary bosonization \cite{gog2}.

In the weak-scattering limit, $\pi V/\omega_{\rm c} \ll 1$,  
the pinning function exhibits more structure. From
 Eq.(\ref{selfcon}) one has
\[
\Omega = V\, (2\pi g V/\omega_{\rm c})^{g/(1-g)}  \;,
\]
which together with Eq.(\ref{cross}) 
implies that the crossover scale goes to infinity as $V\to 0$, namely
$x_0 \sim V^{-1/(1-g)}$. For $x\gg x_0$, SCHA always gives $P\simeq 1$
according to Eq.(\ref{Plarge}). 
This failure is due to the complete neglect of interwell tunneling
in the SCHA, as can be seen by considering the $x\to \infty$ value
of the pinning function (\ref{pinning}),
$ P_\infty = - \langle \cos \bar{q} \rangle_q$, 
where $\bar q$ is the time average value of the 
imaginary-time path $q(\tau)$.
Without tunneling transitions $\bar q$ is an odd
multiple of $\pi$ and one finds $P_\infty=1$
as predicted by SCHA. However, taking into account 
excursions to neighboring wells, it is readily seen that in general
$P_\infty < 1$. 
Despite of these shortcomings, the effective Gaussian treatment 
indicates that for weak scatterers there is a crossover, 
with a slower decay of the Friedel
oscillation at intermediate distances than the asymptotic $x^{-g}$
decay. In fact, Eq.(\ref{respin}) predicts $P \sim x^g$ for
$x\ll x_0$. 

To investigate the weak-scattering limit further, we have evaluated
Eq.(\ref{pinning}) in powers of $V$ giving to lowest order
\begin{equation} \label{pert}
P(x,g,V) = \gamma_g \, \frac{\pi V}{\omega_{\rm c}} 
\left(\frac{x}{\alpha}\right)^{1-g} + {\cal O}(V^3)   
\end{equation}
with 
$\gamma_g = ( 4^{g-1}/\pi) \, B(1/2,g-1/2)$,
where $B(x,y)$ is the Beta function\cite{abram}.
This perturbative result is only valid for $g>1/2$ (otherwise $\gamma_g$
diverges). Furthermore, since higher orders of the perturbation
series grow faster $\sim x^{n(1-g)}$ with $n=3,5,\ldots$, 
the lowest-order result (\ref{pert}) is only valid for $x\ll x_0$
where $x_0$ is found to be given by the SCHA crossover scale (\ref{cross}).
Hence, for intermediate distance from the barrier, the Friedel 
oscillation decays {\em slower} than $x^{-g}$, namely  
like $x^{1-2g}$. As a consequence,
there is a nontrivial limit for the pinning function as $x\to
\infty$ and $V\to 0$. 

\vbox{
\epsfysize=7cm
\epsffile{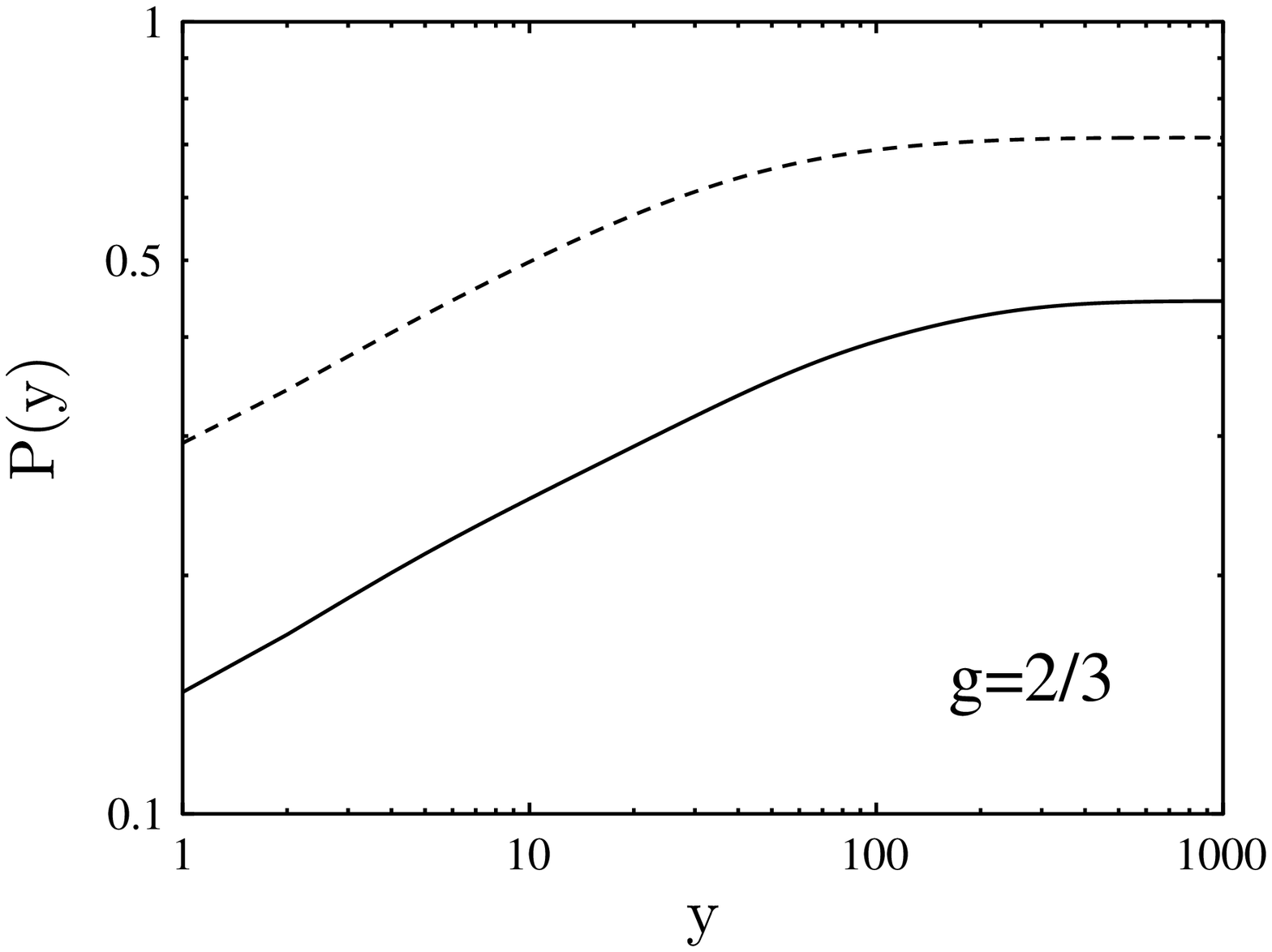}
\begin{figure}
\caption[]{\label{fig1} QMC results for the pinning 
function $P(y)$ at $g=2/3$ and two barrier heights:
 $V/\omega_{\rm c}=0.05$ (solid curve) and 0.1
(dashed curve). The dimensionless
space variable is $y=\omega_{\rm c} x/v_{\rm F}$.
Note the logarithmic scales.}
\end{figure}}

The behavior of the pinning function for arbitrary $V/\omega_{\rm c}$
and $g$  can be computed by means of QMC simulations. In Fig.~1, we show 
results for $g=2/3$ and two (relatively small) barrier heights $V$.
For small-to-intermediate $x$, our data  display
a power law $P\sim x^{\delta_g}$ with $\delta_{2/3} = 0.24\pm 0.03$.
This is in crude accordance with the perturbational result
$\delta_g=1-g$ valid for weak interactions.
On the other hand, for  $x\gg x_0$, the pinning function $P$ is essentially
constant, and the asymptotic decay of the Friedel oscillation is
therefore $\sim x^{-g}$.  A similar behavior is found at $g=1/3$
where direct perturbation theory is inapplicable. Fig.~2 shows QMC results
for the pinning function. The small-to-intermediate $x$ behavior 
is again a power law, now with exponent $\delta_{1/3} = 0.22\pm 0.03$. 
This is in crude accordance with the SCHA prediction $\delta_g=g$
which holds for very strong interactions.
Based on our 
numerical data, the exponents $\delta_g$ are independent of the barrier
height while the region where the intermediate decay is seen shrinks 
rapidly as  $V$ grows.

\vbox{
\epsfysize=7cm
\epsffile{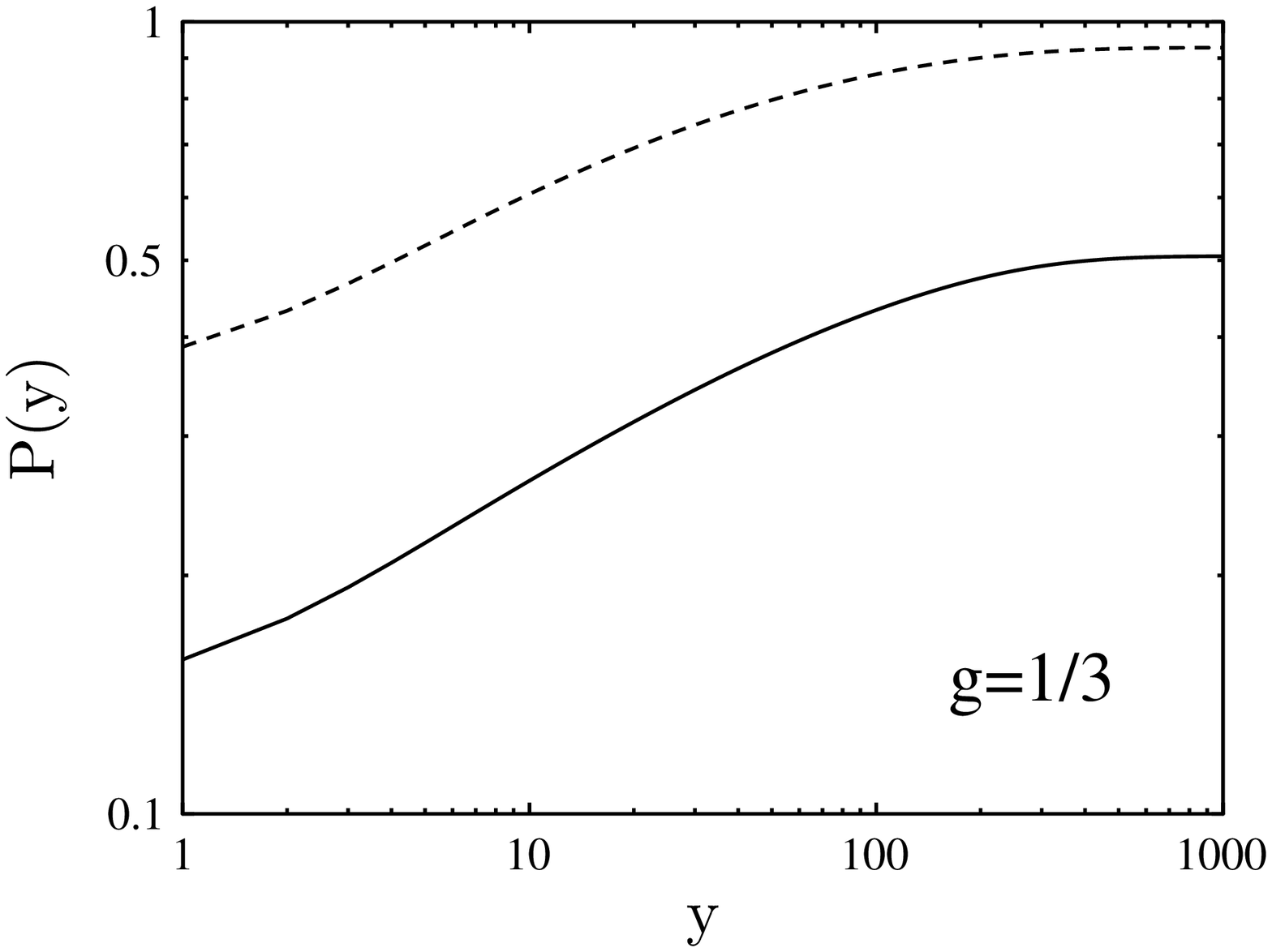}
\begin{figure}
\caption[]{\label{fig2} QMC results for the pinning 
function $P(y)$ at $g=1/3$ and two barrier heights:
 $V/\omega_{\rm c}=0.01$ (solid curve) and 0.05 
(dashed curve).} 
\end{figure}}

For spin-$\frac12$ electrons, we find asymptotically a
slightly faster decay $\sim x^{-(1+g)/2}$ at zero temperature.
This can be rationalized by noticing that
the additional spin channel has $g_\sigma=1$ and the 
exponent $-g$ in Eq.(\ref{friedel}) has to be replaced
by $-(g_\sigma+g_\rho)/2$. 
Remarkably, for spin-$\frac12$ electrons,
there is also a $4 k_{\rm F}$ Friedel oscillation component
\[
\langle \hat \rho (x) \rangle \sim \cos(4k_{\rm F}x)\,x^{-2g} \;, 
\]
which dominates over the $2 k_{\rm F}$ contribution for  
strong enough correlations, $g<1/3$. Since $4 k_{\rm F}$ corresponds 
to the interparticle spacing, this suggests that for $g<1/3$
signatures of Wigner crystal behavior are induced by the
impurity.  

Wigner crystal behavior has also been found by Schulz\cite{wigner}
for the clean system with long-ranged $1/r$ correlations.
For $1/r$ interactions, the $4 k_{\rm F}$ Friedel 
oscillation decay is extremely slow. While the spin  
degrees of freedom involve again the $x^{-1/2}$ factor
suppressing the $2k_{\rm F}$ component, the  
$4 k_{\rm F}$ Friedel oscillations decay like $\exp(-c \sqrt{\ln x})$,
i.e.~slower than any power law.
Effectively, one will then only observe the $4k_{\rm F}$ 
component. 
In the spinless case,  the same quasi long-ranged 
behavior appears for the $2 k_{\rm F}$ component already because the
spin channel is absent now.

Apparently, Friedel oscillations are always present in 1D for 
arbitrary electron-electron interaction. 
Moreover, due to reduced screening in low dimensions,
their decay is always slower than the Fermi liquid $1/x$ prediction. 
We wish to stress that inclusion of backscattering is not expected to
alter these findings substantially. In the spinless case, backscattering
is treated as exchange event of forward scattering and can be 
absorbed by a redefinition of $g$. 
The asymptotic decay of the Friedel oscillation is then always $x^{-g}$.
In the spin-$\frac12$ case, 
based on the renormalization group analysis\cite{solyom},
there are at most weak logarithmic corrections.

To conclude, we have computed the Friedel oscillations in an
interacting 1D electron liquid. 
These results should show up in NMR experiments or 
 as strong quasi-Bragg peaks in x-ray scattering.
They are also of relevance for studies of quasi-one-dimensional 
conductors at low doping concentrations,
or the case of a magnetic impurity.

We wish to thank P. Riseborough, M. Sassetti and U. Weiss
for useful  discussions.



\begin{references}
\bibitem {foot} It is crucial to view $\omega_{\rm c}$
as a physically meaningful bandwidth parameter instead of a regulator
which has to be sent to infinity at any prize.
\bibitem{fot1} We note that the $n=0$ infrared divergence of Eq.(\ref{ff})
cancels in Eq.(\ref{ww}).
\end{references}
\end{document}